\documentclass[sigconf, nonacm]{acmart}

\AtBeginDocument{%
  \providecommand\BibTeX{{%
    \normalfont B\kern-0.5em{\scshape i\kern-0.25em b}\kern-0.8em\TeX}}}

\setcopyright{none}
\renewcommand\footnotetextcopyrightpermission[1]{} 
\begin{document}

\title[Educational Opportunities and Challenges of AI Code Generation]{Programming Is Hard -- Or at Least It Used to Be: Educational Opportunities And Challenges of AI Code Generation}


\author[Becker]{Brett A. Becker}
\orcid{0000-0003-1446-647X}
\affiliation{
  \institution{University College Dublin}
  \city{Dublin}
  \country{Ireland}
}
\email{brett.becker@ucd.ie}

\author[Denny]{Paul Denny}
\orcid{0000-0002-5150-9806}
\affiliation{
  \institution{The University of Auckland}
  \city{Auckland}
  \country{New Zealand}
}
\email{paul@cs.auckland.ac.nz}

\author[Finnie-Ansley]{James Finnie-Ansley}
\orcid{0000-0002-4279-6284}
\affiliation{
  \institution{The University of Auckland}
  \city{Auckland}
  \country{New Zealand}
}
\email{james.finnie-ansley@auckland.ac.nz}

\author[Luxton-Reilly]{Andrew Luxton-Reilly}
\orcid{0000-0001-8269-2909}
\affiliation{
  \institution{The University of Auckland}
  \city{Auckland}
  \country{New Zealand}
}
\email{a.luxton-reilly@auckland.ac.nz}

\author[Prather]{James Prather}
\orcid{0000-0003-2807-6042}
\affiliation{
  \institution{Abilene Christian University}
  \city{Abilene, Texas}
  \country{USA}
}
\email{james.prather@acu.edu}

\author[Santos]{Eddie Antonio Santos}
\orcid{0000-0001-5337-715X}
\affiliation{
  \institution{University College Dublin}
  \city{Dublin}
  \country{Ireland}
}
\email{eddie.santos@ucdconnect.ie}


\begin{abstract}
The introductory programming sequence has been the focus of much research in computing education. The recent advent of several viable and freely-available AI-driven code generation tools present several immediate opportunities and challenges in this domain. In this position paper we argue that the community needs to act quickly in deciding what possible opportunities can and should be leveraged and how, while also working on how to overcome or otherwise mitigate the possible challenges. Assuming that the effectiveness and proliferation of these tools will continue to progress rapidly, without quick, deliberate, and concerted efforts, educators will lose advantage in helping shape what opportunities come to be, and what challenges will endure. With this paper we aim to seed this discussion within the computing education community. 
\end{abstract}

\ccsdesc[500]{Social and professional topics~Computing education}
\ccsdesc[500]{Social and professional topics~Computer science education}
\ccsdesc[500]{Social and professional topics~CS1}
\ccsdesc[500]{Computing methodologies~Artificial intelligence}

\keywords{AI; AlphaCode; Amazon; artificial intelligence; code generation; CodeWhisperer; Codex; Copilot; CS1; CS2; GitHub; Google; GPT-3; introductory programming; machine learning; Midjourney; novice programmers; OpenAI; programming; Tabnine}

\begin{teaserfigure}
  \includegraphics[width=\textwidth]{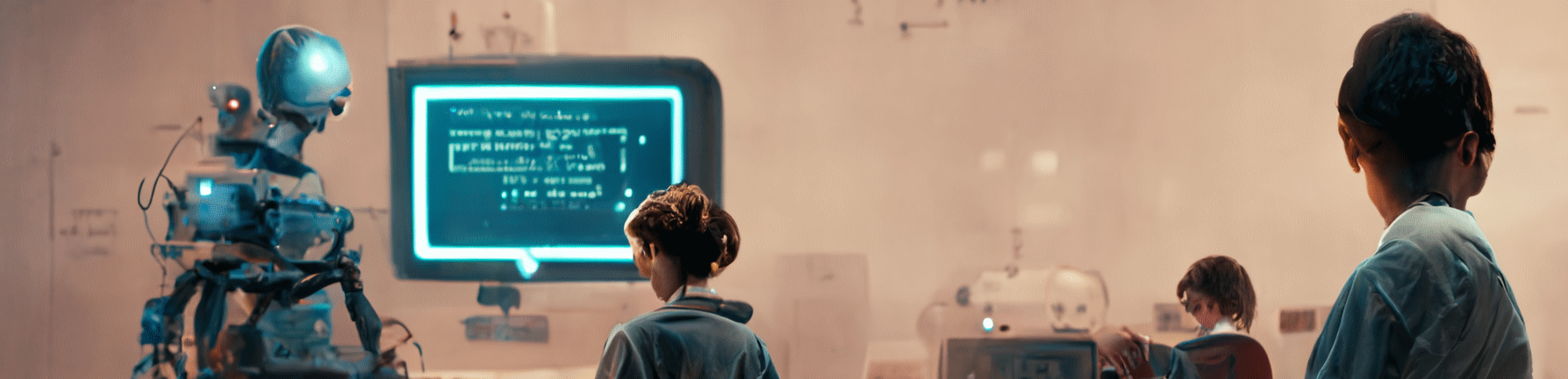}
  \caption{An image generated by Midjourney with the prompt ``robot writing computer code while student watches, computer screens, computer programming, computer code, realistic, highly detailed, cinematic --aspect 16:9''}
  \Description{An illustration of a classroom, with school-aged children sitting behind desks. A robot appears in the place of a teacher, in front of a screen demonstrating program source code. Indecipherable code and diagrams are scattered across the whiteboard in the background.}
  \label{fig:teaser}
\end{teaserfigure}

\maketitle

%
%

\section{Introduction}

Recent months have seen the release of several AI models that represent step-changes in their respective domains. Text-to-image models such as OpenAI's DALL-E 2~\cite{ramesh2022hierarchical} and Midjourney\footnote{
\href{https://www.midjourney.com/home/}{midjourney.com/home}
} (see Figure \ref{fig:teaser}) are revolutionizing how images are created, with the latter being called ``the greatest artistic tool ever built, or a harbinger of doom for entire creative industries''~\cite{blain_2022}. In July 2022, it was announced that DeepMind's AlphaFold predicted the structure of nearly all 200 million proteins known to science and is making them freely available~\cite{heikkila_2022}. Also in the last year, OpenAI and DeepMind -- among others -- have released groundbreaking models that generate computer code.  The model for use so far is that most of these tools will cost money to use professionally but often be free for educational use and to students \cite{wiggers_2022}. It is safe to assume that some computing students are already using AI code completion to generate large chunks of code that could be used in various ways during the completion of assignments.

The introductory programming sequence has been the focus of much research over several decades~\cite{becker201950, luxton-reilly2018introductory} and the challenges of programming at the level required of a first-year computing student have been debated extensively~\cite{becker2021what, luxton-reilly-2016-easy, becker2019compiler}. One particular sticking point is that students should gain extensive practice writing code through dozens of small exercises checked against automated assessment tools \cite{towell2010walls,pettit2017automated}. However students face numerous barriers.  

With the recent advent of several viable AI-driven code generation tools, `writing' code that will suffice to pass traditional first-year programming assignments and even exams seems to have become much easier~\cite{finnieansley2022robots, sarsa2022automatic}.

What does an introductory computing course look like when we can assume that students will be able to easily auto-generate code solutions to their lab and assignment tasks by merely pasting problem descriptions into an AI-powered tool? Further, how will this affect the delivery of computing curricula in general? 
Our view is that these tools stand to change how programming is taught and learned -- potentially significantly -- in the near-term, and that they present multiple opportunities and challenges that warrant immediate discussion as we adapt to the use of these tools proliferating. 

%
%

\section{Background and Context}
Although long-sought, viable AI-driven code generation has only become a reality for the general public since 2021. Three large-scale AI-powered code generation tools -- OpenAI Codex, DeepMind AlphaCode, and Amazon CodeWhisperer -- have recently come into the spotlight. The advantages such systems claim to provide include making ``programming more productive and accessible''~\cite{li2022alphacode}. They also bring several challenges. 

\subsection{OpenAI Codex}
\label{codex}
In September 2021 the \textit{New York Times} published an article titled ``A.I. Can Now Write Its Own Computer Code. That's Good News for Humans'' describing OpenAI's\footnote{\href{https://www.openai.com}{openai.com}} Codex model~\cite{metz2021}. OpenAI is a non-profit ``AI research and deployment company''\footnote{\href{https://openai.com/about/}{openai.com/about/}} set up in 2015 with a \$1 billion pledge from several tech leaders and investors~\cite{shead_2021}. Codex is a descendant of GPT-3 (third-generation Generative Pre-trained Transformer) which is one of the most advanced natural language models available today. 
To demonstrate the kind of output that GPT-3 can generate from a simple prompt, we asked GPT-3\footnote{text-davinci-002 model} the following (after the outline of this paper was determined): \textit{Write a paragraph on the educational opportunities and challenges presented by automated code generation tools.} Its response was:

\begin{quote}There are both educational opportunities and challenges presented by automated code generation tools. On the one hand, these tools can help students learn to code more quickly and efficiently. On the other hand, they can also lead to a false sense of understanding and proficiency, as students may become reliant on the tools to do the heavy lifting for them. Additionally, automated code generation tools can sometimes produce code that is difficult to read and understand, making it more challenging for students to debug and troubleshoot their programs.
\end{quote}

\noindent It is worth noting that this single paragraph from GPT-3 includes some of the opportunities and challenges (in relation to AI-generated code) that we address in this paper.  

Codex is trained on more than 50 million GitHub repositories including the vast majority of GitHub Python code, totaling 159\,GB. Files deemed as likely to be auto-generated, those with an average line count greater than 100, those with maximum line length greater than 1000, and those containing a small percentage of alphanumeric characters were filtered~\cite{chen2021evaluating}. Codex can take English-language prompts and generate code in several languages including JavaScript, Go, Perl, PHP, Ruby, Swift, TypeScript, and shell, but is ``most capable'' in Python\footnote{\href{https://openai.com/blog/openai-codex}{openai.com/blog/openai-codex}}. It can also translate code between programming languages, explain (in English) the functionality of code provided as input, and return the time complexity of code it generates. It also has the ability to generate code that uses APIs, allowing it to, for example, send emails and access information in various databases. Codex is available via the OpenAI API\footnote{\href{https://beta.openai.com/}{beta.openai.com}} and also powers GitHub Copilot\footnote{\href{https://copilot.github.com/}{copilot.github.com}} which is billed as ``Your AI pair programmer'' -- an intentional reference to pair programming, a well-known software engineering~\cite{beck2000extreme} and programming education approach~\cite{mcdowell2002effects}.  Copilot is now available for free to verified students and teachers.\footnote{\href{https://github.com/pricing}{github.com/pricing}} 

The Codex model has been shown to perform well when solving programming tasks presented in plain English. 
The paper announcing Codex solved 29\% of the problems in a new evaluation set developed by the Codex authors to measure functional correctness for synthesizing programs from Python docstrings. This performance increased to 70\% when repeated sampling is employed~\cite{chen2021evaluating}.  

The first evaluation of Codex on introductory programming problems was reported by Finnie-Ansley et al.~\cite{finnieansley2022robots}, who compared its performance on summative exam questions to that of students in an introductory course, and found that it outperformed almost 80\% of the students in the course.  In addition, it comfortably solved various definitions of the classic Rainfall Problem~\cite{soloway1986learning}, including one novel variation that had never been published.

\subsection{DeepMind AlphaCode}
In February 2022, DeepMind\footnote{\href{https://www.deepmind.com/}{deepmind.com}} announced AlphaCode\footnote{\href{https://alphacode.deepmind.com/}{alphacode.deepmind.com}} which, like Codex, utilizes a transformer-based model that ``writes computer programs at a competitive level''\footnote{\href{https://www.deepmind.com/blog/competitive-programming-with-alphacode}{deepmind.com/blog/competitive-programming-with-alphacode}}. It is trained on over 715\,GB of GitHub code including programs written in C++, C\#, Go, Java, JavaScript, Lua, PHP, Python, Ruby, Rust, Scala, and TypeScript~\cite{li2022alphacode}. All files larger than 1\,MB or with lines longer than 1000 characters, and duplicates of the same file (ignoring whitespace) were filtered from training data. Unlike Codex, AlphaCode was fine-tuned on a curated set of publicly released competitive programming problems called CodeContests.\footnote{\href{https://github.com/deepmind/code\_contests}{github.com/deepmind/code\_contests}} In introducing AlphaCode, Li et al.\ claim that a stripped-down version of AlphaCode, without the modifications described in their paper, performs similarly to Codex, ``however, problems used in the Codex paper and similar work consist of mostly simple task descriptions with short solutions -- far from the full complexity of real-world programming''~\cite{li2022alphacode}.

AlphaCode ranked in the top 54\% of over 5,000 programming competition participants from the Codeforces platform, solving new problems requiring a combination of critical thinking, logic, algorithms, coding, and natural language understanding~\cite{li2022alphacode}. Based on these results, the authors estimate that AlphaCode has a Codeforces\footnote{\href{https://codeforces.com/}{codeforces.com}} rating of 1238 which is within the top 28\% of users that participated in a contest in the last 6 months~\cite{li2022alphacode}. Li et al.\ also showed that AlphaCode does not duplicate sections of code from the training dataset when producing solutions, instead relying heavily on natural language problem descriptions to create original solutions. AlphaCode is not currently available as an API or otherwise.

\subsection{Amazon CodeWhisperer}
Amazon CodeWhisperer was announced in June 2022~\cite{ankur_atul_2022}. Unsurprisingly a Google Scholar search (July 27, 2022) returned only four results for \textit{amazon codewhisperer} none of which pertain to the tool itself. CodeWhisperer is billed as ``the ML-Powered Coding Companion''\footnote{\href{https://aws.amazon.com/codewhisperer/}{aws.amazon.com/codewhisperer}} which ``helps improve developer productivity by providing code recommendations based on developers' natural comments and prior code''~\cite{ankur_atul_2022}. Based on (for instance) a developer comment describing a task, CodeWhisperer attempts to determine which cloud services and public libraries are best for the task, generates code, and presents this as a recommendation to the developer within the IDE. Like Codex and AlphaCode, it is trained on public data. It is also claimed that accuracy is directly proportional to the size of the training data~\cite{ankur_atul_2022} -- a finding similar to that of Codex~\cite{tamkin2021understanding}. 
CodeWhisperer is currently available for free, subject to a waitlist.\footnote{\href{https://pages.awscloud.com/codewhisperer-sign-up-form.html}{pages.awscloud.com/codewhisperer-sign-up-form.html}} 

\subsection{Other AI code generation products}

Although Codex, AlphaCode, and CodeWhisperer are the most publicized AI-driven code generation platforms, several others exist including  Tabnine\footnote{\href{https://www.tabnine.com/}{tabnine.com}}, Code4Me\footnote{\href{https://code4me.me/}{code4me.me}} and FauxPilot\footnote{\href{https://github.com/moyix/fauxpilot}{github.com/moyix/fauxpilot}} (based on SalesForce CodeGen~\cite{Nijkamp2022ACP}). Most of these tools are commercial offerings aimed at professional software developers, as one of the oft-touted (although currently unproven) advantages of AI-driven code generation is increased development productivity.

\section{Position}

Our position is the following: \textit{AI-generated code presents both opportunities and challenges for students and educators in introductory programming and related courses. 
The sudden viability and ease of access to these tools suggest educators may be caught unaware or unprepared for the significant impact on education practice resulting from AI-generated code.  We therefore urgently need to review our educational practices in the light of these new technologies.
}

We take it as given that these tools will continue to be readily available to students (as they are currently), that adoption will increase, and that the capabilities of the tools will improve. In the following sections we describe some of the opportunities and challenges presented by AI code generating tools in the context of university-level novices learning to program in the present time. We largely focus on opportunities and challenges that are already well-documented in the computing education literature, and discuss how AI-generated code is likely to affect the landscape of areas that are already well-studied. Where available, we include evidence and results from the literature although the literature on the effects of cutting-edge AI code generation tools is in its infancy.  We intend this presentation of opportunities and challenges to form the basis for the inevitable discussions about the role of code generation tools in our education practices.

In this work we do not discuss wider societal (e.g., economic~\cite{chen2021evaluating}, political~\cite{khlaaf2022hazard}) considerations presented by AI-generated code, nor those specific to advanced/professional programmers (e.g.,~\cite{moroz2022potential,dakhel2022github}). While important issues, our  focus is on how AI-generated code is likely to impact students and educators in introductory programming (and related) classrooms in the near term. 

While any new technology brings with it both positive and negative impacts, the hope is always that long-term net effects are positive.  
The developers of Codex specifically note that they do not ``expect the impact of this class of technologies to be net-negative; rather, risks merit particular attention ... because they may be subtle or require deliberate effort to address, whereas [they] expect the benefits to be more obvious and ``automatic'' from the perspective of most users and affected stakeholders''~\cite{chen2021evaluating}.

The challenges and opportunities presented here are not exhaustive, but rather starting points for ongoing discussions that we hope lead to best educational practices involving code generation tools.

\section{Opportunities}

Any new tool, when effective and widely-available, poses opportunities for learning. Handheld calculators have been ubiquitous in mathematics education since the 1980s. A study in (US) grades K-12 statistically analyzing 524 effects from 79 separate studies recommended the use of calculators in all mathematics classes from kindergarten (approx.\ age 5) on, including in testing situations for grades 5 (approx.\ age 10) and up~\cite{hembree1986effects}. 
It remains to be seen if AI-powered code generation will follow a similar path.

Opportunities noted by the developers of Codex include ``the potential to be useful in a range of ways'' including: ``help onboard users to new codebases; reduce context switching for experienced coders; enable non-programmers to write specifications; have [such tools] draft implementations; and aid in education and exploration''~\cite{chen2021evaluating}. The developers of AlphaCode also see obvious opportunities in such tools, suggesting ``the potential for a positive, transformative impact on society, with a wide range of applications including computer science education, developer tooling, and making programming more accessible''~\cite{li2022alphacode}.

In this section we offer a number of avenues where AI-generated code tools present clear opportunities for computer science education. Although some opportunities bring related challenges, in this section we focus on their benefits.

\subsection{Code Solutions for Learning}

\subsubsection{Exemplar solutions}
Students learning to program are typically encouraged to practice writing code by completing short problems.  However, students are sometimes unable to complete the exercises, and even when successful often seek exemplar solutions.  
Unfortunately, instructors do not always have the time to prepare and publish model solutions for all the programming exercises that students engage in (which may include test and exam questions). AI-generated solutions provide a low-cost way for students to generate exemplar solutions to check their work when practicing~\cite{finnieansley2022robots}.

\subsubsection{Variety of solutions}
Code generation tools can also be used to help expose students to the variety of ways that a problem can be solved.  There are usually many different approaches for solving a programming problem, although novices do not always appreciate this.  Thompson et al.\ argue that providing appropriate variation in programming instruction is important, because it helps learners to appreciate the efficiencies and differences in approaches to writing code \cite{thompson2006code}.  
Eckerdal and Thun\'{e} make a similar argument, drawing on variation theory to state that teachers should make available resources that highlight dimensions of variation in concepts being studied \cite{eckerdal2005novice}.  For most non-trivial problems, code generation tools produce a variety of correct solutions that are offered to the programmer for selection.  This was illustrated by Finnie-Ansley et al.\ who observed a great deal of variation in the solutions that Codex generated when solving the classic Rainfall problem~\cite{finnieansley2022robots}.  

\subsubsection{Code review of solutions}

Current assessment approaches in introductory programming courses often focus on code correctness, rather than code quality or style.  With the ability to generate syntactically-correct solutions automatically, assessment can focus on the differences between multiple correct solutions, and making judgments on the style and quality of solutions.  

Extensive literature on peer review, including code reviews \cite{indriasari2020review,luxton-reilly2009systematic}, outline the many benefits from looking at a variety of solutions to a given problem. These benefits are reportedly present even when the code is flawed -- there are benefits from looking at good solutions as well as poor ones.  Code generation models could be used to generate solutions of varying, or unknown quality, and these could be used for assessment tasks focusing on the evaluation of code quality to engage students at the highest level of Bloom's taxonomy. 
This may prove useful for generating discussions around alternative approaches and the quality of solutions, and provide the basis for refactoring exercises~\cite{finnieansley2022robots}.

Current models are effective at generating correct code, but to our knowledge, no studies have looked at the style of AI-generated code.  We believe that future models will have more sophisticated methods of selecting high-quality code that adheres to style conventions.
The developers of AlphaCode note that automatic code generation could make programming more accessible and help educate new programmers \cite{li2022alphacode}. These models could suggest alternative, and more efficient or idiomatic ways of implementing programs, which could help learners to improve their coding style.

\subsection{Producing Learning Resources}

Generating high-quality learning resources is very time consuming and typically requires a high level of expertise.  The potential for generating novel learning resources, like programming exercises,  explanations of code, and worked examples at essentially an unlimited scale is an exciting avenue for future work.  

\subsubsection{Exercise generation}
Very recent work by Sarsa et al. has shown that the Codex model is capable of producing novel learning resources from a single priming example \cite{sarsa2022automatic}.  They explored the generation of two types of resources -- programming exercises and code explanations -- finding that most of the generated exercises were sensible and novel and included an appropriate sample solution~\cite{sarsa2022automatic}. They also reported that the Codex model was very effective at producing contextualized problem statements, targeting certain thematic topics that were simply specified as input to the model as part of the priming example.  

\subsubsection{Code explanations}

High quality explanations of code are a useful type of resource for helping learners develop a robust understanding of programming concepts.  One of the widely publicized features of Codex is that it can generate explanations of complicated pieces code.  The example of this functionality provided as part of the OpenAI playground uses the prompt: ``Here's what the above class is doing: 1.'' (with the number at the end prompting the model to produce an enumerated list when describing a code fragment).   In their study of learning resource generation, Sarsa et al.\ found that most of the explanations generated by Codex were thorough and correct~\cite{sarsa2022automatic}. 
Recent work by MacNeil et al. also explored different kinds of prompts and the diverse code explanations they lead to when using the GPT-3 language model~\cite{macneil2022generating}.

\subsubsection{Illustrative examples}
Texts and other learning resources typically provide examples that are used to learn the relationship between a described programming problem and a solution.  These can be used to illustrate a given programming construct, algorithmic pattern, data structure, or mapping from problem to solution.  Students use these examples as models that help them learn, and frequently express a desire for more examples than are available.  Code generation tools provide a means of satisfying this desire and providing as many examples as needed.  
As AI code generation tools improve, this could lead to worked examples that include reasoning for coding decisions, similar to those generated by Minerva for mathematics problems~\cite{lewkowycz-2022-minerva}.  Such examples are believed to lower cognitive load and result in more effective learning~\cite{knorr2019reforming}.

\subsection{New Pedagogical Approaches}
Teaching in CS1 typically focuses initially on syntax and basic programming principles, and it usually takes time for students to master these fundamentals.  If code generation models can be used to solve the low level implementation tasks, this may allow students to focus on higher level algorithms.  In a way, this is similar to the use of block-based environments that remove the complexities of syntax and allow students to focus on algorithmic issues.  Teaching could initially focus more on algorithms and problems solving, relying on automatic code generation for implementation, and then delay discussions of syntax until later.

\subsubsection{Explaining algorithmic concepts clearly}

The way that prompts to code generation models are constructed affects their performance. 
Simplifying the problem description was found to significantly increase the success rate of AlphaCode. On a sample of difficult problems, simplifying the description to make the required algorithm more explicit increased the percentage of correct samples from 12\% to 55\%~\cite{li2022alphacode}.  It was also found that sensitivity to consistent variable naming decreases with model size -- random changes to variable names in problem descriptions mattered less. That is, models are  increasingly able to capture relevant relationships between variables described in the problem formulation.  Students can focus more on how to communicate algorithmic problems clearly, thereby providing a better description to code generation models that can then generate working solutions.

\subsubsection{Alleviating programmer's writer's block}

Anecdotally, students sometimes struggle with programmer's writer's block -- that is, they don't know how to get started. Vai\-tha\-lingam et al.~\cite{vaithilingam2022expectation} found that Copilot helped students to get started with programming assignments by producing some starter code, thereby offering the opportunity to \emph{extend} code rather than struggling with a blank page. Such an approach may require us to shift focus towards rewriting, refactoring, and debugging code; however, this provides the opportunity to help students maintain forward momentum in an authentic environment where the need for evaluating, rewriting, and extending code is perhaps more important than writing every line of code from scratch~\cite{minelli2015know}.


\subsubsection{Overcoming traditional barriers}

Novices face many barriers in learning to program~\cite{becker201950}. For instance, programming (compiler) error messages are a known barrier to student progress~\cite{becker2019compiler, karvelas2020compiler}. Recent work has demonstrated that Codex is capable of explaining error messages in natural language - often effectively - and that that it can also provide correct fixes based on input code and error messages~\cite{leinonen2023using}. It is likely that the efficacy of these approaches will improve in time, and that other barriers to novice learning may be similarly mitigated by such models.

\section{Challenges}
\label{challenges}The availability AI-based code generation raise concerns that it could be used in ways that limit learning, or in ways that make the work of educators more difficult.
The developers of Codex note that their tool ``raises significant safety challenges,
does not always produce code that is aligned with user intent, and has the potential to be misused''~\cite{chen2021evaluating}. Similarly, the developers of AlphaCode note that ``like most technologies, these models might enable applications with societal harms which we need to guard against, and desire to have a positive impact is not itself a mitigation against harm''~\cite{li2022alphacode}. In this section we present a number of challenges presented by AI-generated code tools. Although some of these may also present opportunities, in this section we focus on their challenges. 

\subsection{Ethical Issues}

Academic integrity in computing is a complex issue, particularly when software development encourages reuse of code and collaborative practices~\cite{simon.ea-2016-maze}. The use of auto-generated code raises significant issues with respect to academic integrity and code reuse.

\subsubsection{Academic misconduct}
Prior work has shown that AI-generated code tools can achieve better than average marks on actual student exams, can perform well on both standard programming questions such as Rainfall~\cite{finnieansley2022robots}, and reliably generate correct code for common algorithms such as insertion sort and tree traversal~\cite{dakhel2022github}. We can assume that AI-generated code tools will be \textit{capable} of completing assignments that we give to students learning programming.  
    
Simon et al.~\cite{simon.ea-2016-maze} note that contract cheating is growing in prevalence and increasingly difficult to detect.  This suggests that there is student \textit{desire} to outsource graded work to others.  Traditional outsourced solutions have risks that communication between student and provider may be breached, or that the solution may be shared (or reused) by the provider resulting in duplicate submissions that can be detected.  AI-generated solutions vary~\cite{finnieansley2022robots}, and do not require communicating with another person, producing similar results to contracted outsourcing for students with fewer inherent risks. This provides a low-risk/high-reward avenue for students focused on short-term grades rather than developing a deep understanding of content.  This may exacerbate existing issues with detection of academic misconduct.

\subsubsection{Attribution}
Simon et al.~\cite{simon.ea-2016-maze} surveyed academics about the use of attribution for code obtained from outside sources, finding a diverse range of views on the acceptability of code reuse.  This academic integrity quagmire becomes more complex with relatively opaque differences between standard code completion tools present in IDEs and plugins such as Copilot that will provide code suggestions that are indistinguishable from IDE code completion.

In other contexts, we use spell-checkers, grammar-checking tools that suggest rewording, predictive text and email auto-reply suggestions -- all machine-generated.  In a programming context, most development environments support code completion that suggests machine-generated code. Distinguishing between different forms of machine suggestions may be challenging for academics, and it is unclear if we can reasonably expect introductory programming students who are unfamiliar with tool support to distinguish between different forms of machine-generated code suggestions.  If students are unable to distinguish these, then it would be unjust to treat use of machine-generated code as academic misconduct.  This raises a key philosophical issue: how much content can be machine-generated while still attributing the intellectual ownership to a human?  This calls into question the very concept of plagiarism~\cite{dehouche2021plagiarism} and how we should interpret plagiarism and intellectual contribution with machine-supported generation of content.

\subsubsection{Code reuse and licensing}
There are also potential licensing issues that arise when new content is produced using code generation models, even when the model data is publicly-available~\cite{li2022alphacode}.  Many different licenses apply to much of the publicly-available code and typically these require authors to credit the code they used, even when the code is open-source.  When the use of that code comes about via an AI model, developers may end up using code that requires license compliance and not be aware that it does.\footnote{\href{https://githubcopilotinvestigation.com/}{githubcopilotinvestigation.com/}} This is clearly an issue that extends beyond educational use of software, but as educators it is our role to inform students of their professional responsibilities when reusing code.

\subsubsection{Sustainability}
The sustainability of our education practices, and in particular the impact on our environment, is an ethical issue that we must acknowledge. Training AI models can consume significant energy.~\citet{brown2020language} report that GPT-3/Codex required more than several thousand petaflop/s-days of computation during the pre-training process.  Further, due size these models are hosted centrally and accessed remotely. At present these tools are likely not as efficient in terms of compute power and network traffic than more established web services and their environmental costs are a sustainability concern that should be known to those using them.

\subsection{Bias and Bad Habits}
The issue of bias in AI is well known~\cite{bender2019typology}. In addition to general bias (subtle or overt) that applies to almost all AI-generated outputs such as only representing certain groups of people, genders, etc., there are likely biases specific to AI code generation.  

\subsubsection{Appropriateness for beginners}
Given that most of the code that these models are trained on is public, it is reasonable to question if the public code used for training is appropriate for students who are starting to learn programming.  For example, professionals (and the over-confident) are likely more amenable to posting their code publicly. This can be used to support an argument that, despite myriad examples that public code is not very good, it is nonetheless on average of higher quality than non-public code. At least most public code is complete and is subject to public scrutiny -- something that can not be said for private code. In addition to this quality bias, the code styles of public code is likely different -- and possibly more advanced than that of a typical ``blank slate'' novice. However, these styles and approaches may not match those of the instructor. 

\subsubsection{Harmful biases}
The developers of Codex note found that code generation models raise bias and representation issues beyond problematic natural language -- notably that Codex can generate code with structure that reflects stereotypes about gender, race, emotion, class, the structure of names, and other characteristics~\cite{chen2021evaluating}.  Additionally, Codex ``can be prompted in ways that generate
racist, denigratory, and otherwise harmful outputs as code comments''~\cite{chen2021evaluating}.

\subsubsection{Security}
Although largely ignored for much of the short history of computing education, the requirement for novices programmers to begin learning secure coding practices has been well-documented in recent years~\cite{lam2022identifying}. Given this, the security of AI-generated code is extremely important, even in educational settings. It has been shown that code generated by these models can be insecure~\cite{pearce2022asleep}, and human oversight is required for the safe use of AI code generation systems ~\cite{chen2021evaluating}.  CodeWhisperer claims to tackle security head-on by providing the ability to run scans on code 
to detect security vulnerabilities~\cite{ankur_atul_2022}, although this is currently untested. Chen et al. noted that although future code generation models may be able to be trained to produce more secure code than the average developer, this is far from certain~\cite{chen2021evaluating}.

\subsection{Over-reliance}

The Codex developers noted that a key risk of using code generation models in practice is users' over-reliance on the generated outputs~\cite{chen2021evaluating}. Novices using such models, especially with tools such as Copilot that embed support in an IDE, may quickly become accustomed to auto-suggested solutions.  This may lead to students not reading problem statements carefully, or at all, and therefore not thinking about the computational steps needed to solve a problem.

\subsubsection{Reinforcing behaviors that reduce learning}
An analysis of solutions generated by AlphaCode revealed that 11\% of Python solutions were syntactically-incorrect (produced a \texttt{SyntaxError}) and 35\% of C++ solutions did not compile~\cite{li2022alphacode}. It is not known what the average compilation rate of submitted solutions for average introductory programming students studying these languages are, however it is clear that students using AlphaCode and other AI-generated code tools would be dealing with code that has a high probability of being incorrect in some way. 
\balance
The Codex developers noted that it can recommend syntactically-incorrect code including variables, functions, and attributes that are undefined or outside the scope of the codebase.  
\citet{chen2021evaluating} observe ``Codex may suggest solutions that superficially appear correct but do not actually perform the task the user intended. This could particularly affect novice programmers, and could have significant safety implications depending on the context.''  If suggested code is incorrect, students may lose trust in the feedback provided by IDEs, including error messages, warnings and other auto-generated forms of feedback.

\section{Conclusions}
AI-generated code is now firmly part of the education landscape, but we do not yet know how to adapt our practices to overcome the challenges and leverage the benefits.  What we confidently predict is that software development of the future will include an increasing amount of auto-generated code and this includes those training for such roles and jobs, such as our students.  We believe this minimally suggests a shift in emphasis towards code reading and evaluating rather than code generation -- a pedagogical approach consistent with the theory of instruction advanced by \citet{xie.ea-2019-instruction}.  Beyond pedagogy, it also demands we examine the ethical implications of the use of these tools and that we guide our students through such ethical reflection.  In a 2022 ITiCSE keynote, Titus Winters, a principal software engineer at Google, suggested it's at least as important to be an ethically-aware person as it is be a good programmer \cite{winters2022gap}.  We believe AI-generated code coupled with demands from industry will force us to face ethical issues in computing education from the very beginning of the curriculum. 
Without quick, concerted efforts, educators will lose advantage in helping shape what opportunities come to be, and what challenges will endure.

\bibliographystyle{ACM-Reference-Format}
\bibliography{sample-base}


\end{document}